\documentclass{emulateapj}
\usepackage{amsmath}
\usepackage{amsfonts} 
\usepackage{appendix}

\usepackage{float}

\shorttitle{How dead are dead galaxies?}
\shortauthors{Mattia Fumagalli et al.}
\slugcomment{ApJ, accepted} 

\begin{document}
\title{How dead are dead galaxies? Mid-Infrared fluxes of quiescent galaxies at redshift $0.3 < z < 2.5$: implications for star-formation rates and dust heating.}

\author{Mattia Fumagalli\altaffilmark{1}, Ivo Labb{\'e}\altaffilmark{1}, Shannon G. Patel\altaffilmark{1},  Marijn Franx\altaffilmark{1}, Pieter van Dokkum\altaffilmark{2}, Gabriel Brammer\altaffilmark{3}, Elisabete da Cunha\altaffilmark{4},  Natascha M. F\"{o}rster Schreiber\altaffilmark{5}, Mariska Kriek\altaffilmark{6}, Ryan Quadri\altaffilmark{7}, Hans-Walter Rix\altaffilmark{4}, David Wake\altaffilmark{8}\altaffilmark{13}, Katherine E. Whitaker\altaffilmark{9}, Britt Lundgren\altaffilmark{8}, Danilo Marchesini\altaffilmark{10}, Michael Maseda\altaffilmark{4}, Ivelina Momcheva\altaffilmark{2}, Erica Nelson\altaffilmark{2}, Camilla Pacifici\altaffilmark{11}, Rosalind E. Skelton\altaffilmark{12}}

\altaffiltext{1}{Leiden Observatory, Leiden University, P.O. Box 9513, 2300 RA Leiden, Netherlands}
\altaffiltext{2}{Department of Astronomy, Yale University, New Haven, CT 06511, USA}
\altaffiltext{3}{European Southern Observatory, Alonso de Córdova 3107,Casilla 19001, Vitacura, Santiago, Chile}
\altaffiltext{4}{Max Planck Institute for Astronomy (MPIA), Königstuhl 17, 69117 Heidelberg, Germany}
\altaffiltext{5}{Max-Planck-Institut für extraterrestrische Physik, Giessenbachstrasse,D-85748 Garching, Germany}
\altaffiltext{6}{Department of Astronomy, University of California, Berkeley, CA 94720, USA}
\altaffiltext{7}{Observatories of the Carnegie Institution of Washington, Pasadena, CA 91101, USA}
\altaffiltext{8}{Department of Astronomy, University of Wisconsin, Madison, WI 53706, USA}
\altaffiltext{9}{Astrophysics Science Division, Goddard Space Flight Center, Code 665, Greenbelt, MD 20771, USA}
\altaffiltext{10}{Department of Physics and Astronomy, Tufts University, Medford, MA 02155, USA}
\altaffiltext{11}{Yonsei University Observatory, Yonsei University, Seoul 120-749, Republic of Korea}
\altaffiltext{12}{South African Astronomical Observatory, Observatory Road, Cape Town, South Africa}
\altaffiltext{13}{The Open University, Department of Physical Sciences, The Open University, Milton Keynes, MK7 6AA, UK}

\begin{abstract}

We investigate star-formation rates (SFR) of quiescent galaxies at high redshift ($0.3 < z < 2.5$) using 3D-HST WFC3 grism spectroscopy and {\it Spitzer} mid-infrared data. We select quiescent galaxies on the basis of the widely used UVJ color-color criteria. 
Spectral energy distribution (SED) fitting (rest-frame optical and near-IR) indicates very low star-formation rates for quiescent galaxies ($\rm sSFR \sim 10^{-12} yr^{-1}$). However, SED fitting can miss star formation if it is hidden behind high dust obscuration and ionizing radiation is re-emitted in the mid-infrared. It is therefore fundamental to measure the dust-obscured SFRs with a mid-IR indicator.
We stack the MIPS-24$\mu$m images of quiescent objects in five redshift bins centered on $z$ = 0.5, 0.9, 1.2, 1.7, 2.2 and perform aperture photometry. Including direct 24$\mu$m detections, 
we find $\rm sSFR \sim 10^{-11.9}$ $\times (1+z)^{4} yr^{-1}$.
These values are higher than those indicated by SED fitting, but at each redshift they are 20-40 times lower than those of typical star-forming galaxies. 
The true SFRs of quiescent galaxies might be even lower, as we show that the mid-IR fluxes can be due to processes unrelated to ongoing star formation, such as cirrus dust heated by old stellar populations and circumstellar dust. 
Our measurements show that star-formation quenching is very efficient at every redshift. The measured SFR values are at $z > 1.5$ marginally consistent with the ones expected from gas recycling (assuming that mass loss from evolved stars refuels star formation) and well below that at lower redshifts.

\end{abstract}


\section{Introduction}

A bimodal distribution in galaxy properties (star-formation rate, size, morphology) has been observed in the local Universe (e.g. Kauffmann et al. 2003). This bimodality is made of blue, predominantly late-type galaxies, whose emission is dominated by young stellar populations and experiencing significant level of star formation, complemented by red, early-type (elliptical or S0) galaxies dominated by an old stellar population with little or absent star formation.

The bimodality has been observed all the way to z $\sim$ 2 (Labb{\'e}  et al. 2005, Kriek et al. 2006, Ilbert et al. 2010, Brammer et al. 2011, Whitaker et al. 2013). 

Specific star-formation rates (sSFR) from spectral energy distribution (SED) fitting and equivalent widths from emission lines indicate for quiescent galaxies very low values ($\rm log_{10} sSFR \cdot yr ~< -12$) even at high redshift (Ciambur et al. 2013, Kriek et al. 2006, Whitaker et al. 2013), suggesting that these galaxies really are dead.

These levels of star formation are much lower than expected. Even if the galaxy would have stopped accreting new gas from the intergalactic medium, some gas should always become available again for star formation due to gas recycled from evolved stages of stellar evolution (e.g. Leitner \& Kravtsov, 2010). If the low levels of star formation are confirmed, it could have important implications for gas recycling and the effectiveness of quenching at high redshift. Alternatively, it is possible that amounts of star formation have been overlooked in previous studies because of heavy obscuration by dust. 

\begin{figure*}[!t!h]
\centering
\includegraphics[width=19.2cm]{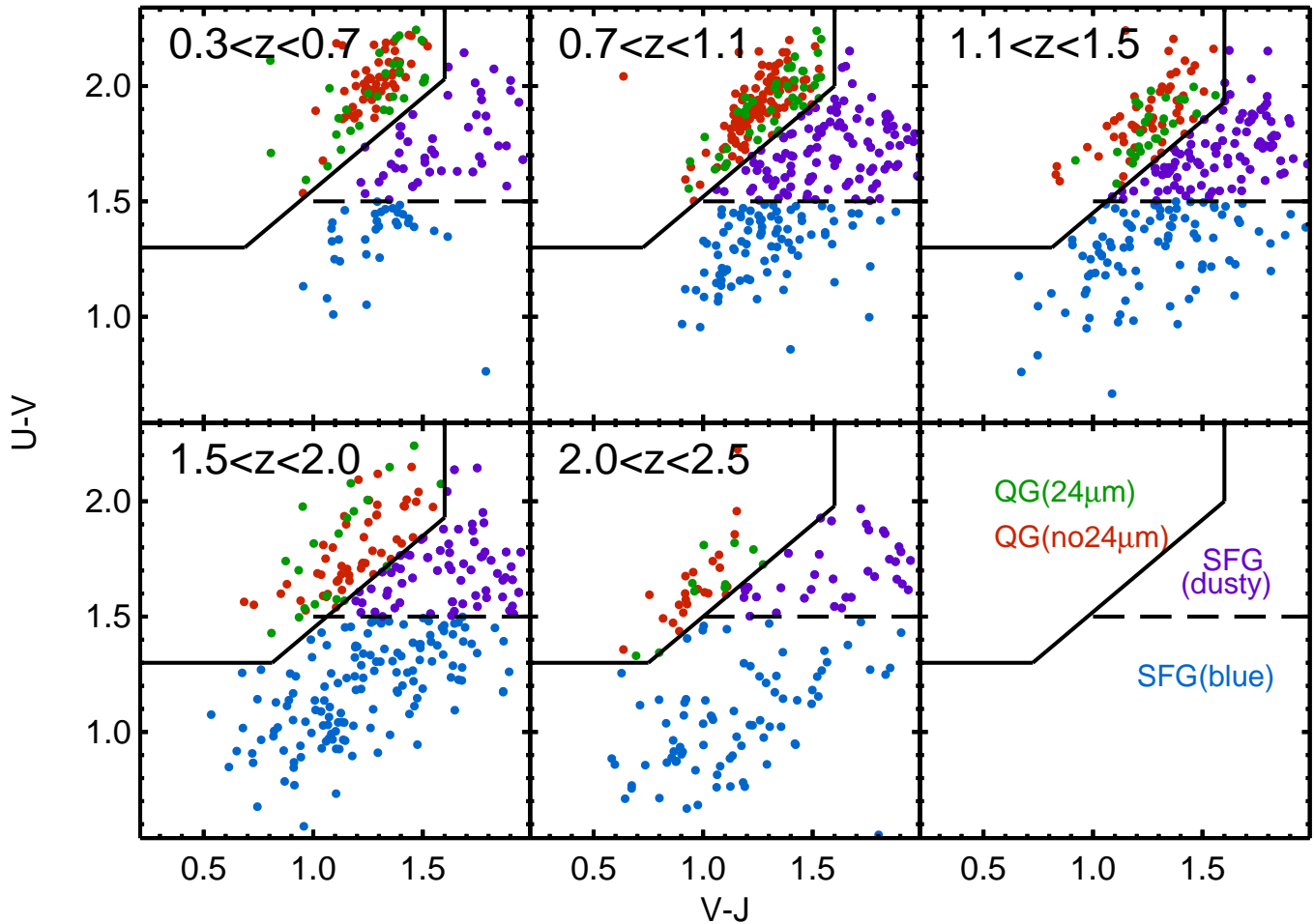}
\caption{UVJ selection in different redshift bins, for mass-selected samples (${\rm log_{10}(}M_{\star}/M_{\odot}{\rm )>10.3}$). The Whitaker et al. (2012) boundary divides (solid black line) quiescent and star-forming galaxies. SFGs are subdivided into dusty ($U-V > 1.5$, purple dots) and unobscured ($U-V < 1.5$, blue dots). QGs are color coded according to the presence of a 24$\mu$m detection. We notice that 24$\mu$m-detected galaxies do not preferentially lie in a particular locus of the UVJ diagram. }
\label{UVJ}
\end{figure*}

\begin{figure*}[!htb!]
\centering
\includegraphics[width=13.5cm]{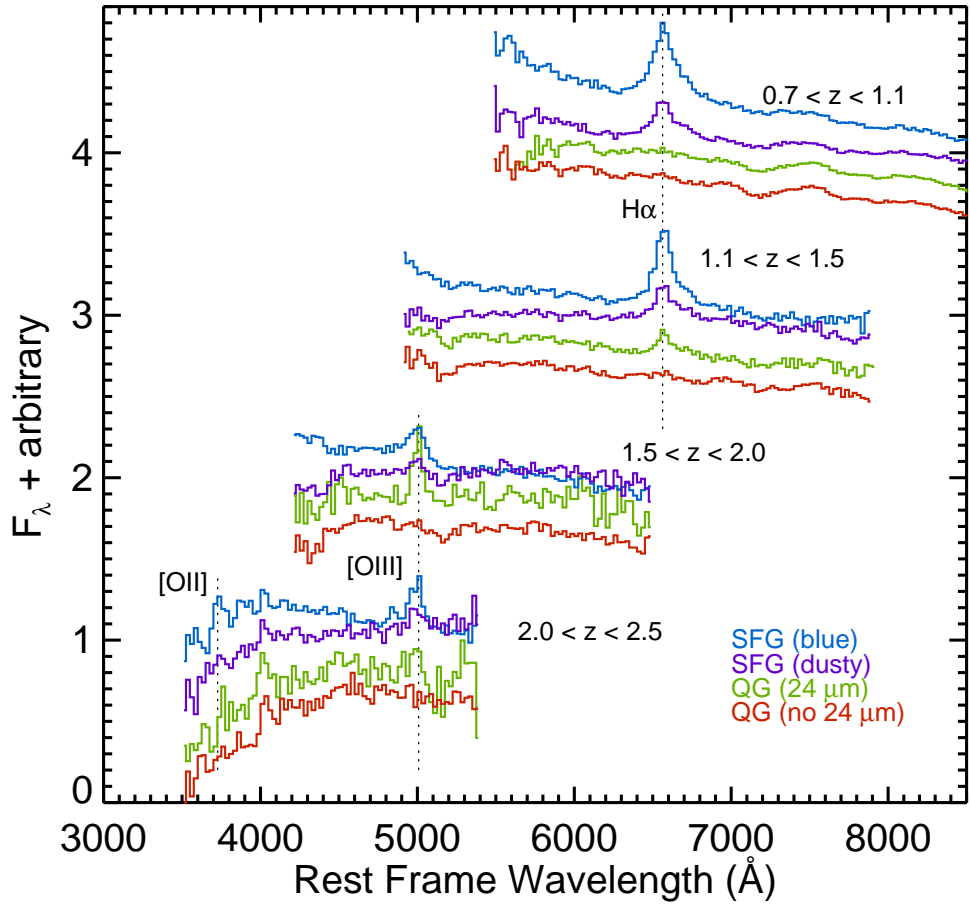}											     	  
\caption{Stacked 3D-HST spectra for mass-selected (${\rm log_{10}(}M_{\star}/M_{\odot}{\rm )>10.3}$) galaxies in different redshift bins. In each redshift bin, blue means blue SFGs ($U-V < 1.5$), purple dust-reddened SFGs ($U-V > $ 1.5), green QGs with a 24$\mu$m detection, red QGs without a 24$\mu$m detection. Vertical dashed lines show the position of H$\alpha$, [OIII] and [OII].} 
\label{spectrastackplot2}
\end{figure*}

\begin{figure*}[!t!h]
\centering
\includegraphics[width=14.3cm]{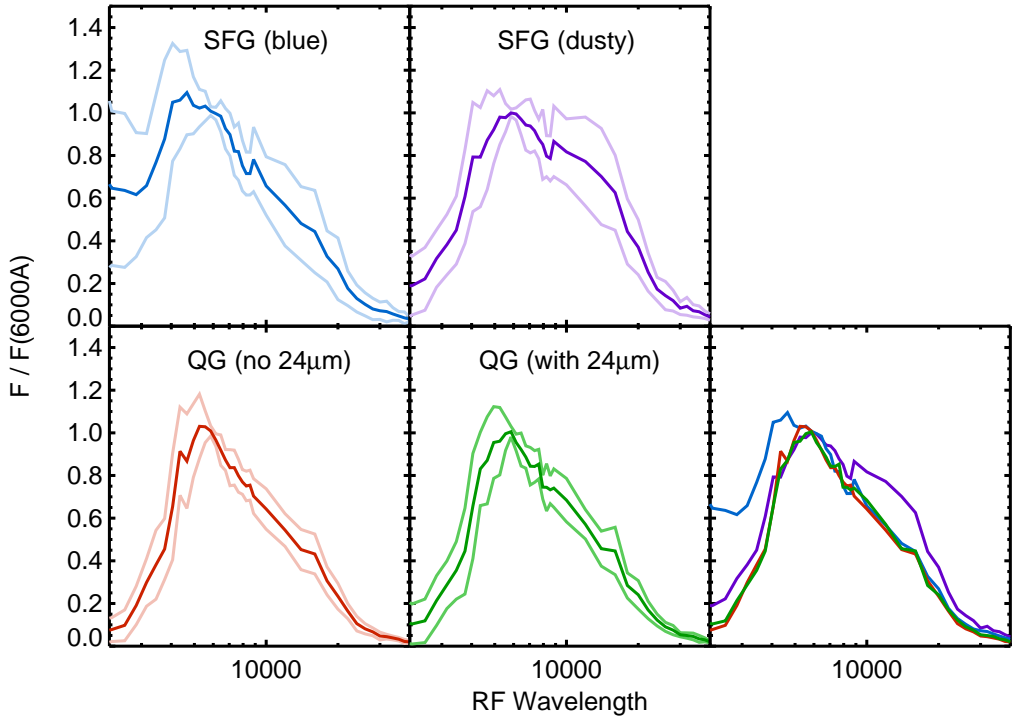}
\caption{Composite SEDs for mass-selected samples (${\rm log_{10}(}M_{\star}/M_{\odot}{\rm )>10.3}$) of star-forming galaxies (divided into blue and dusty) and quiescent galaxies (divided according to the presence of a 24$\mu$m detection) at redshift $0.3 < z < 2.5$. Light lines indicate the scatter in the stacks. In the bottom-right panel we overplot the four composite SEDs, showing that quiescent galaxies with and without 24$\mu$m detection have very similar optical and near-IR SED shapes, while star-forming galaxies and dusty star-forming galaxies are clearly distinguishable.}
\label{composite_SED}
\end{figure*}

\begin{figure}[!h!]
\centering
\includegraphics[width=8.cm]{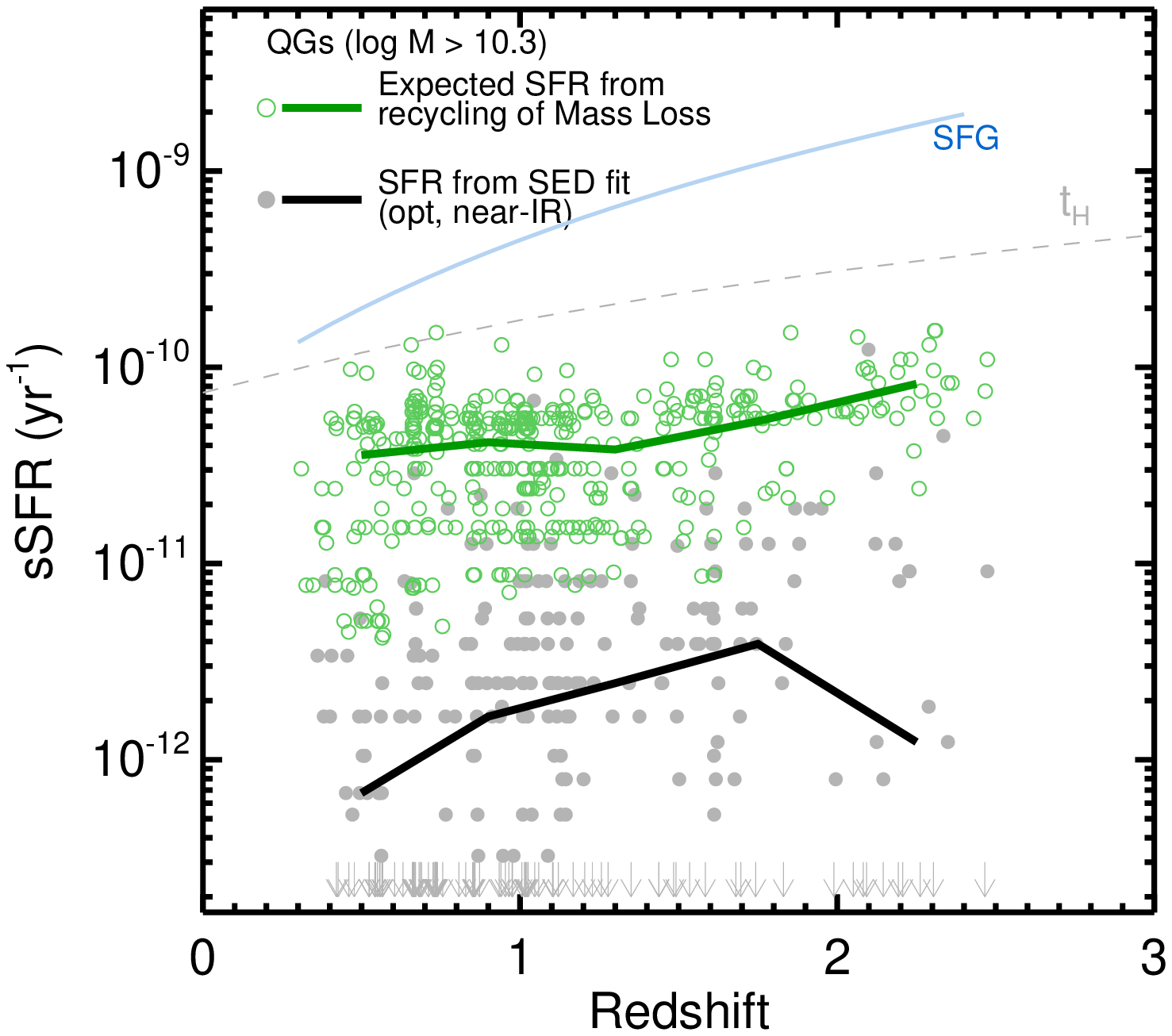}
\caption{
sSFR in different redshift bins (gray filled dots) and expected sSFR from recycling of gas from the mass loss of evolved stars (green open dots), as determined from FAST best fits to the SEDs of quiescent galaxies. Solid lines represent mean values in different redshift bins. The mass loss is computed from $M_{gas}$ in BC03 models. It overpredicts the SFR by a factor of 20 at each redshift. sSFRs of star-forming galaxies on the main sequence (cyan) and the Hubble time (dashed gray) are shown as references. 
}
\label{FASTvsMassLoss}
\end{figure}

To address this question, in this paper we determine the obscured SFRs of quiescent galaxies up to redshift $z \sim 2.5$ using their 24$\mu$m emission. 
In Section 2, we discuss the data. In Section 3, we describe the selection of QGs, and compare their SFRs from optical and near-infrared (IR) SED fitting to the values expected from the recycling of gas from mass loss. We additionally evaluate how much obscured star formation might be hidden in our selection: this proves the need of looking at a mid-IR indicator for SFR. In Section 4, we stack 24$\mu$m thumbnails of QGs in order to measure their obscured SFR. We evaluate possible contributions to the mid-IR fluxes of QGs in Section 5. We discuss our findings in Section 6 and summarize them in Section 7. Through the paper we assume a standard cosmology with $\rm H_{0} = 70 km \, s^{-1} \, Mpc^{-1}$, $\rm \Omega_{M} = 0.30$, and $\rm \Omega_{\Lambda} = 0.70$ (Komatsu et al. 2011).

\section{Data}

The 3D-HST Survey (van Dokkum et al. 2011; Brammer et al. 2012) is a 600 arcmin$^2$ survey using the Hubble Space Telescope (HST) to obtain complete, unbiased low-resolution near-IR spectra for thousands of galaxies. (Cycles 18 and 19, PI: van Dokkum).

\begin{figure}[!h!]
\centering
\includegraphics[width=8cm]{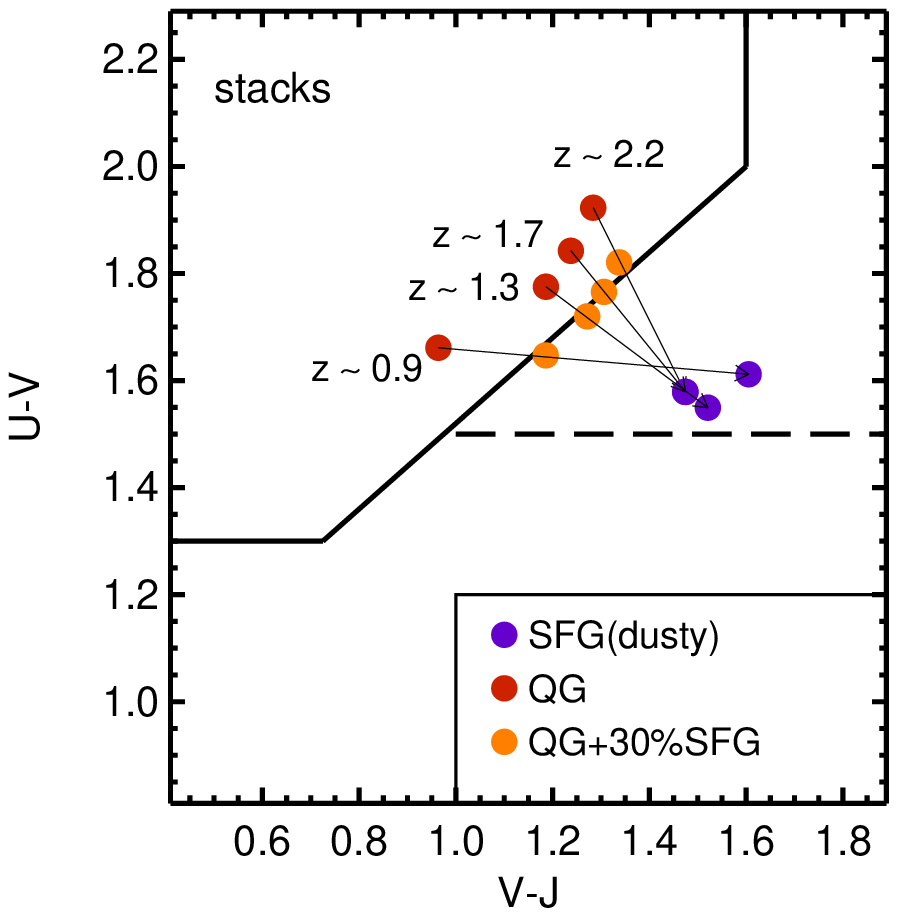}
\caption{The UVJ position of stacks of QGs (red) and dusty SFGs (purple) at different redshifts is shown. Black arrows show the tracks obtained summing a variable fraction (normalized at 6000 $\rm \AA$, $\rm F_{SFG}$) of the dusty SFGs SEDs to the QGs SEDs. Orange dots show composite SEDs on the UVJ boundary, corresponding to $\rm F_{SFG} = 30$ \%.}
\label{UVJpercentage}
\end{figure}

It targets five fields (COSMOS, GOODS-S, GOODS-N\footnote{GOODS-N has been taken as part of program GO-11600 (PI:
B. Weiner) and integrated into 3D-HST}, AEGIS, UDS) where a wealth of ancillary multi-wavelength data are available ($U$ band to 24$\mu$m); they are crucial for interpreting spectra that often contain a single emission line, if any. 
The 3D-HST photometric catalogue is described in detail in Skelton et al. (2014). It contains $\sim170000$ sources, detected on a noise-equalized combination of the F125W, F140W and F160W images. The completeness of 3D-HST as a function of magnitude is evaluated by comparing the number of detections in the catalog to those in a deeper image of GOODS-S: the two of them deviate at magnitudes fainter than $F160 = 25$ mag.

The WFC3 grism spectra have been extracted with a custom pipeline, described in Momcheva et al. (2014, in prep). Redshifts have been measured via the combined photometric and spectroscopic information using a modiﬁed version of the EAZY code (Brammer et al. 2008). The precision of redshifts is proven to be $\sigma(\frac{dz}{1+z})=0.3\%$ (Brammer et al. 2012, Momcheva et al. 2013). 

Accurate redshifts allow the derivation of accurate rest-frame fluxes: we interpolate rest-frame filters from the observed SED with the Inter-rest code (Taylor et al., 2009), based on the algorithm by Rudnick et al (2003). Stellar masses have been determined using the FAST code by Kriek et al. (2009), using Bruzual \& Charlot (2003) models, and assuming exponentially declining star-formation histories (with e-folding times $log_{10}(\tau/yr)$ ranging from $10^{7}$ to $10^{11}$ yr), solar metallicity and a Chabrier (2003) IMF. 
The 3D-HST catalogs are evaluated to be 90$\%$ complete in stellar mass down to ${\rm log_{10}(}M_{\star}/M_{\odot}{\rm)>9.4}$ at $z<2.5$ (Tal et al. 2014).

In this paper we restrict the analysis to the GOODS-N and GOODS-S fields, for which very deep {\it Spitzer}-MIPS  ($S_{24\mu m}= {\rm 10 \mu Jy}$, 3$\sigma$) data are available (Dickinson et al., 2003), necessary for inferring low levels of SF. The MIPS 24$\mu$m beam has a FWHM of 6 arcsec, therefore confusion and blending effects are unavoidable in deep observations at this resolution. 
We perform photometry using a source-fitting algorithm (Labb{\'e} et al. 2006, Wuyts et al. 2007) that takes advantage of the higher resolution information contained in the F160W images, as described in the Appendix. This method produces a model PSF for the image with the broader native PSF (MIPS-24$\mu$m in our case), which is then used to estimate the flux distribution of each source identified in the detection image ($F160W$) segmentation map output by SExtractor. For each individual object, the flux from neighboring sources (closer than 10 arcsec) is fitted and subtracted, allowing for a reliable aperture flux measurement of individual objects.

Total IR luminosities ($L_{{\rm IR}}$ = $L({\rm 8-1000 \mu m})$) were derived from the observed 24$\mu$m fluxes, on the basis of a single template that is the average of Dale \& Helou (2002) templates with 1 $<$ $\alpha$ $<$ 2.5, following Wuyts et al. (2008; see also Franx et al. 2008, Muzzin et al. 2010), and in good agreement with recent {\it Herschel}/PACS measurements by Wuyts et al. (2011). 
SFRs are determined from the IR emission as in Bell et al. (2005) for a Chabrier IMF:
${\rm SFR(IR) = 0.98 \times 10^{-10} }L_{{\rm IR}}(L_{\odot})$.\footnote{
The Bell et al.(2005) relation properly applies to starbursts of continuous star formation, with recent star-formation timescale of $\sim 10^8$yr and solar metallicity.} 
This quantity accounts properly just for obscured SF. 

We derive SFRs without using data at wavelengths longer than 24$\mu$m because photometry from the PACS and SPIRE instruments on {\it Herschel} in the GOODS fields is not as deep as that from MIPS-24$\mu$m. We evaluate the potential for detecting low SFRs with the {\it Herschel} instruments, by using the PACS and SPIRE detection limits (Elbaz et al. 2011) to SFRs: we extrapolate total IR luminosities from monochromatic fluxes, and convert them to SFR as described above. We find that at $z=1$ the PACS-100$\mu$m and 160$\mu$m photometry is able to detect SFRs higher than $\sim 2M_{\odot}/yr$ (1$\sigma$), and SPIRE-250 higher than $\sim 5M_{\odot}/yr$ (1$\sigma$), while at the same redshift MIPS-24 $\mu$m is one order of magnitude deeper ($\sim 0.3M_{\odot}/yr$). The same conclusion holds true at $z=2$, with detection limits for MIPS-24$\mu$m, PACS-100$\mu$m, PACS-160$\mu$m and SPIRE-250$\mu$m being respectively $\sim 2M_{\odot}/yr$, $\sim 20M_{\odot}/yr$, $\sim 10M_{\odot}/yr$, $\sim 20M_{\odot}/yr$ (all 1$\sigma$ limits).

On the 3D-HST GOODS fields extremely deep X-ray data are also available, 4Ms in CDF-South (see Xue et al. 2011), and 2Ms in CDF-North (see Alexander et al. 2003), that we use for identifying bright AGNs.

\begin{figure*}[!t!]
\centering
\includegraphics[width=18.4cm]{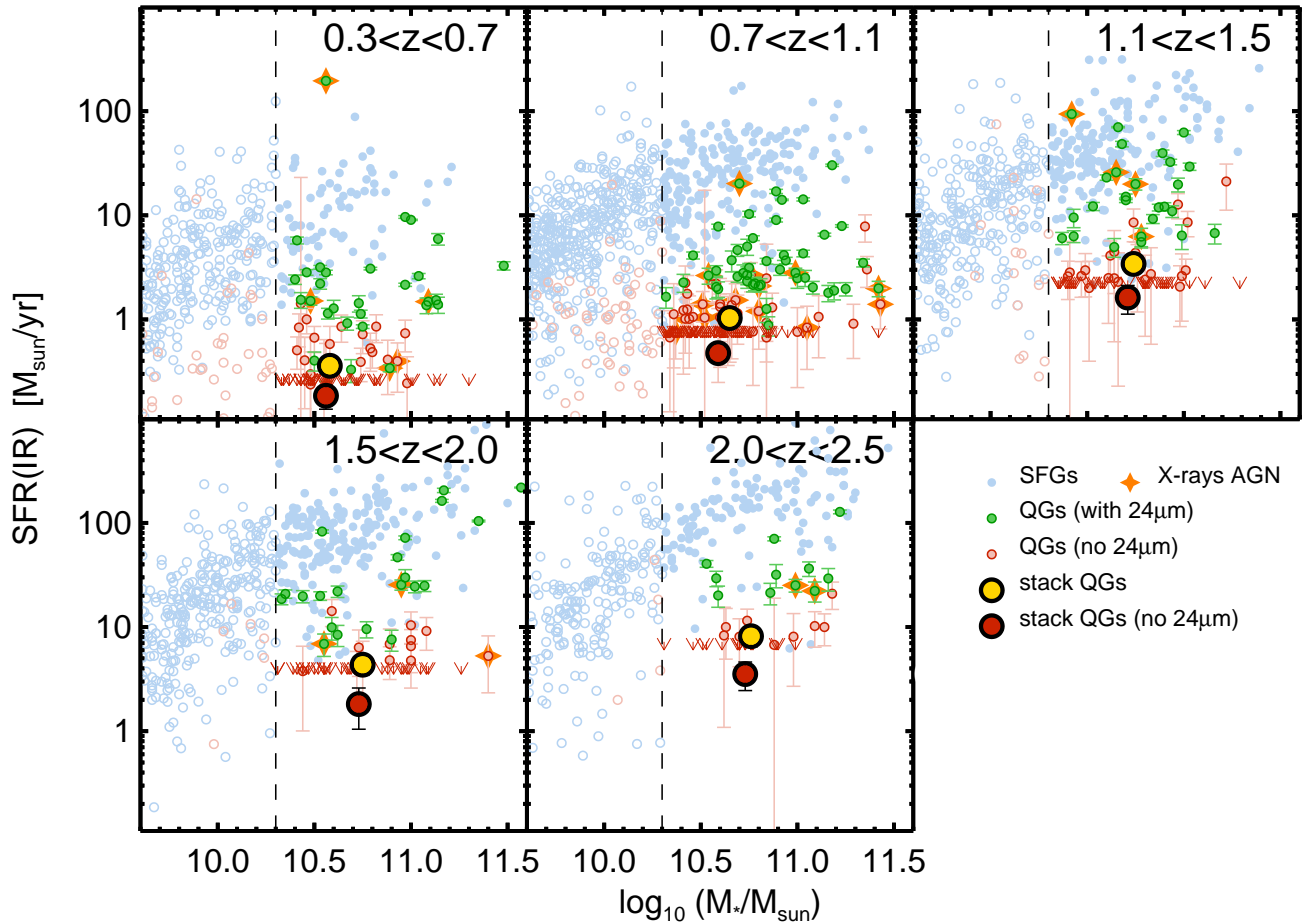}											     	  
\caption{
Mass-SFR(IR) diagram for galaxies in the 5 redshift bins analyzed in the paper. SFRs are computed assuming that all of the 24$\mu$m flux is due to reprocessed UV photons from HII regions. Filled symbols denote galaxies with ${\rm log(}M_{\star}/M_{\odot} {\rm )>10.3}$. Galaxies are divided in quiescent (QGs) and star-forming (SFGs) according to the box defined in Whitaker et al. (2012): light-blue dots represent SFGs, green dots QGs detected at 24$\mu$m ($\rm S/N > 3$), red dots QGs not detected at 24$\mu$m. Sources with $\rm S/N < 1$ are shown as arrows at the $1\sigma$ level. Orange stars represent X-ray detected QGs in the CDF-S 4Ms catalog and CDF-N 2Ms catalog. The large red dots show the SFRs obtained stacking thumbnails of individually undetected QGs (red), and all QGs (yellow), in mass-selected samples of ${\rm log_{10}(}M_{\star}/M_{\odot}{\rm )>10.3}$. Errors on the stacks are computed through bootstrapping of the sample.}
\label{MassSFR}
\end{figure*}

\begin{figure*}[!t!]
\centering
\includegraphics[width=18.4cm]{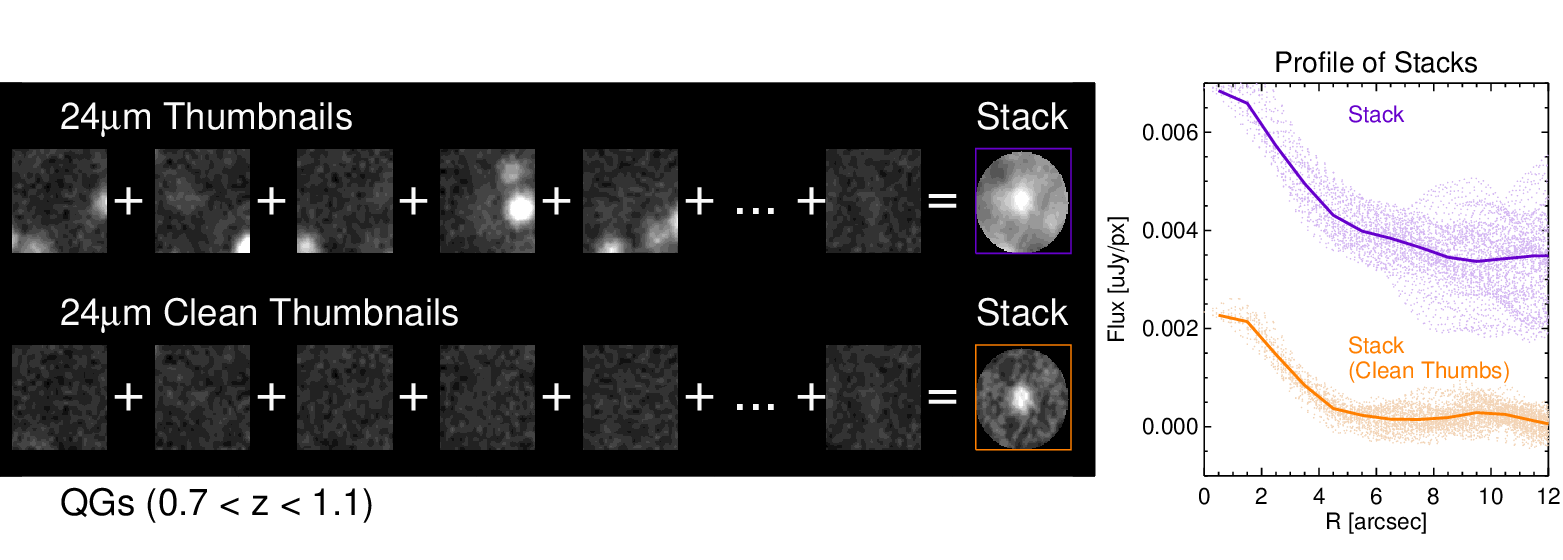}											     	  
\caption{Stacking procedure for quiescent galaxies. Left: we show on the top row a random sample of 24$\mu$m postage stamps (20'' wide) from the QGs sample at redshift $0.7<z<1.1$ and the resulting stack. The same postage stamps after the neighboring sources have been subtracted (with the PSF-matching technique described in Section 2) are shown in the bottom row, along with the resulting mean stack. Right: Profile of the mean stacks (lines) and individual pixel values (dots) at their distance from the center. The 'normal' stack has a high, poorly constrained, background level (artificially raised by neighboring sources).}
\label{stack_procedure}
\end{figure*}

\section{Sample selection and motivations of the study}
\subsection{Selection of Quiescent Galaxies}
In order to select quiescent galaxies (QGs) we use a color-color technique (Figure \ref{UVJ}), specifically rest-frame $U-V$ versus rest-frame $V-J$ (hereafter: UVJ diagram). This technique has been widely used to distinguish QGs from SFGs, including the heavily reddened SFGs (Labb{\'e} et al. 2005; Wuyts et al. 2007; Williams et al. 2009; Brammer et al. 2009; Whitaker et al. 2010; Patel et al. 2012; Bell et al. 2012; Gobat et al. 2013). 
QGs are identified using the criteria $(U-V) > 0.8 \times (V-J)+0.7$, $U-V > 1.3$ and $V-J < 1.5$, as in Whitaker  et al. (2012) \footnote{We test the stability of the selection by shifting the box by $\pm 0.05$ magnitude, which does not affect the analysis.}.
Effectively, this selection targets galaxies whose optical and near-IR light is dominated by an old stellar population. We select galaxies more massive than ${\rm log_{10}(}M_{\star}/M_{\odot}{\rm )>10.3}$ and divide the sample in five redshift bins, centered on $z$ = 0.5, 0.9, 1.2, 1.7, 2.2. The sample is mass complete at $>97.5 \%$ level even at the highest redshift ($z<2.5$) we consider (Tal et al. 2014).
At each redshift the QG sample consists of at least 60 galaxies (Table 1).

\subsection{Spectra and SEDs of the sample}
In Figure \ref{spectrastackplot2} we show stacked optical spectra of QGs and SFGs from 3DHST in mass-selected samples (${\rm log_{10}(}M_{\star}/M_{\odot}{\rm )>10.3}$). SFGs are subdivided into blue SFGs ($U-V <$ 1.5) and dust-reddened SFGs ($U-V >$ 1.5). 

QGs are subdivided according to the presence of a MIPS 24$\mu$m detection. As noted by other authors (e.g. Brammer et al. 2009, Barro et al. 2013), approximately $25\%$ of the optically-selected QGs have a 24$\mu$m detection, which is in apparent contrast with the red optical colors and the SEDs. We also notice that 24$\mu$m-detected QGs do not lie preferentially in any locus of the UVJ diagram (green dots in Figure 1). 

The spectra in Figure \ref{spectrastackplot2} clearly show that the UVJ selection is efficient in dividing the two populations; the SFG selection includes the heavily dust-reddened SFGs, that despite red $U-V$ colors, show spectral features (H$\alpha$, D4000) characteristic of SFGs. It is also noteworthy to see that QGs with 24$\mu$m detections have some H$\alpha$ and [OIII] (cfr. Whitaker et al. 2013), that indicate the presence of low level star formation and/or nuclear activity. 

Figure \ref{composite_SED} shows composite SEDs (following the methodology of Kriek et al. 2011) for SFGs (divided into blue and dusty) and QGs (divided according to the presence of a 24$\mu$m detection). The SED shapes of star-forming galaxies, dusty star-forming galaxies and quiescent galaxies are clearly different. The rest-frame optical and near-IR SEDs of QGs with and without 24$\mu$m detection are instead very similar. 

\subsection{SFRs from SED fitting and expectations from gas recycling}

We first analyze the sSFR from the SED fits to the UV-optical and near-IR photometry
(see Section 2). The values are shown against redshift for the quiescent galaxies in Figure 4 (gray dots and black line). The median value is $\rm sSFR = 1.7 \times 10^{-12} yr^{-1}$, and the correlation with redshift is weak. These values compare well with those of Ciambur et al (2013), who used a similar method.
No significative difference in the SED-derived SFRs is seen if we split the quiescent population between galaxies with and without a 24$\mu$m detection.

The low sSFRs can be compared to the stellar mass loss from evolved stellar populations (Parriott \& Bregmann, 2008;  Leitner \& Kravtsov, 2010).
Green dots in Figure \ref{FASTvsMassLoss} represent for the QG sample the sSFR expected from stellar mass loss, assuming that 100\% of the gas expelled from old stars is recycled into star formation. 
Mass loss is computed directly from $\rm M_{gas}$ of the BC03 models at the best fit age of the galaxy, given the best fit $\tau$ model (see Section 2, Data). 
The expected sSFR from gas recycling is $\rm 2-4 \times 10^{-11} yr^{-1}$, with a weak redshift dependence. It overpredicts the sSFR from optical and near-IR SED fitting by more than one order of magnitude.

The discrepancy between the two values at each redshift tells us that one of the following options must hold true: 
\begin{itemize}
\item a mechanism able to prevent the cooling of gas expelled from old stars and therefore the fueling of new star formation exists, {\it or}
\item SFRs from optical and near-IR best fitting are underestimated (and a lot of star formation shows up in the mid-IR).
\end{itemize}
In the rest of the paper we test the latter possibility measuring SFRs in the mid-IR, in order to prove the former.

\subsection{How much star formation could be hidden?}
We evaluate how much star formation a galaxy can hide (with high dust obscuration), while still retaining red optical-NIR colors. We stack the rest-frame SEDs of QGs and dusty SFGs in different redshift bins; to each QG SED we add a variable fraction ($\rm F_{SFG}$, normalized in light at 6000$\rm \AA$) of the dusty SFG SED. Figure \ref{UVJpercentage} shows the position in the UVJ diagram of the QGs stacks (red), the dusty SFGs stacks (purple) and the SED with $\rm F_{SFG} = 30$\% (orange), on the UVJ separation border. Adding a 30\% dusty SFG SED to our typical QG SED would keep such a galaxy as quiescent under our selection criteria despite the non-negligible contribution of obscured star formation.  Since the SFR of the average SFG evolves with redshift, $\rm F_{SFG} = 30$\% corresponds to $\rm sSFR \sim 8 \times 10^{-11} yr^{-1}$ at $z=0.5$ and $\rm sSFR \sim 3 \times 10^{-10} yr^{-1}$ at $z=2.2$. This shows that with high dust content, a red (optical and near-IR) galaxy can hide a significant amount of SF. It is therefore necessary to measure SFR from MIR indicators in order to evaluate the SFRs of QGs. 
There is also a potential for entirely obscured populations with $A_{V} {\rm \gg 5}$, which are known to exist the the centers of local dusty star-bursting galaxies (e.g. Arp 220, Sturm et al. 1996).

\section{Measuring Obscured Star-Formation Rates of Quiescent Galaxies}

In this Section we discuss the SFRs determined from the IR emission with the methodology described in Section 2.
In Figure \ref{MassSFR} we plot the relation between stellar mass and SFR for galaxies in the mass-selected sample. As already noticed by various authors using a variety of SFR indicators (e.g. Noeske et al. 2007, Damen et al. 2009, Whitaker et al. 2012), SFRs and masses of SFGs are correlated (light-blue dots), with a scatter of approximately 0.3 dex. The vast majority of the QGs lies in this plane below the 'star-forming main sequence'. Most of the QGs are undetected in the MIPS 24$\mu$m image at 3$\sigma$ (red dots), while some of them (approximately 25 \%, Table 1) have a 24$\mu$m detection (green dots), placing them in the Mass-SFR plane between SFGs and the detection limit. 

In order to measure the SFRs for QGs, we stack 24$\mu$m thumbnails. We emphasize that in this step we stack \emph{cleaned} images (20'' wide), where neighboring sources identified in the high resolution $F160W$ band have been subtracted with the technique described in the Appendix.

Summing original 24$\mu$m thumbnails would lead to a stack with a very poorly constrained background, raised by the presence of neighboring objects. Since the goal of this paper is to measure very low SFRs with accuracy, it is fundamental to perform photometry on a stacked image with small uncertainty on the background (as shown in Figure \ref{stack_procedure}).

We perform an average-stacking\footnote{Using instead median stacks does not modify the conclusions of the paper.} in different redshift bins, for two samples: all QGs and only non-24$\mu$m-detected QGs. Photometry on the stack is performed within an aperture of 6 arcsec diameter, similar to the size of the 24$\mu$m FWHM. To measure the total 24$\mu$m flux, we create a MIPS growth curve from several bright, isolated, and unsaturated point sources within each field. These square postage stamps are 20 arcsec wide, and we derive an aperture correction of a factor of 2.57 from r =3 arcsec to r = 10 arcsec. To convert to total flux, we include an additional aperture correction for the 22\% of the flux that falls outside 10 arcsec, derived from the MIPS handbook. \footnote{Since the calibration for MIPS refers to an object with $\rm T=10000K$, we color-correct fluxes by dividing them by a factor of 0.967 (MIPS Handbook, Table 4.17). This way flux densities correspond to those of sources with a flat spectrum.}

\begin{figure*}[!t!h]
\centering
\includegraphics[width=8.5cm]{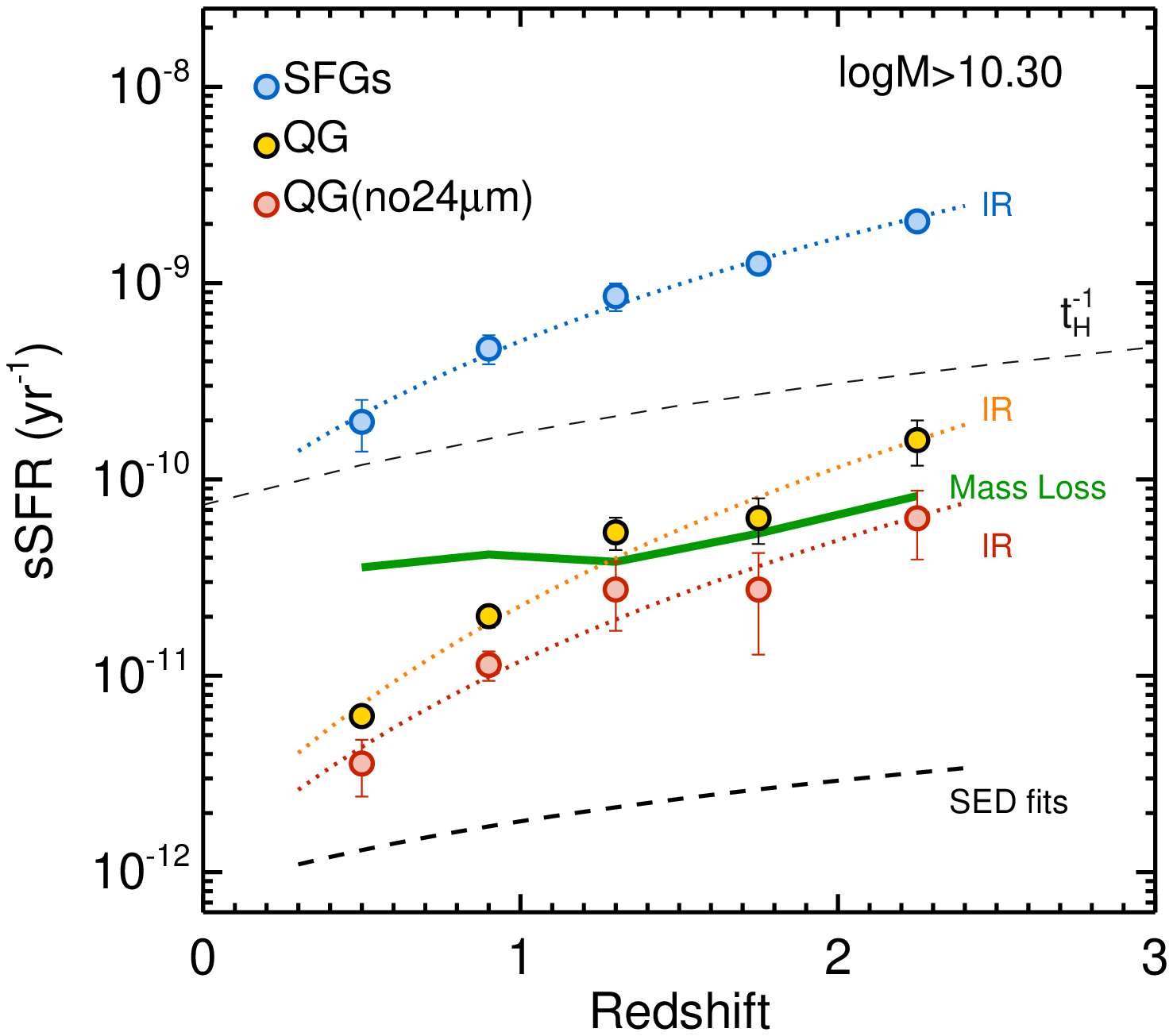}
\includegraphics[width=8.5cm]{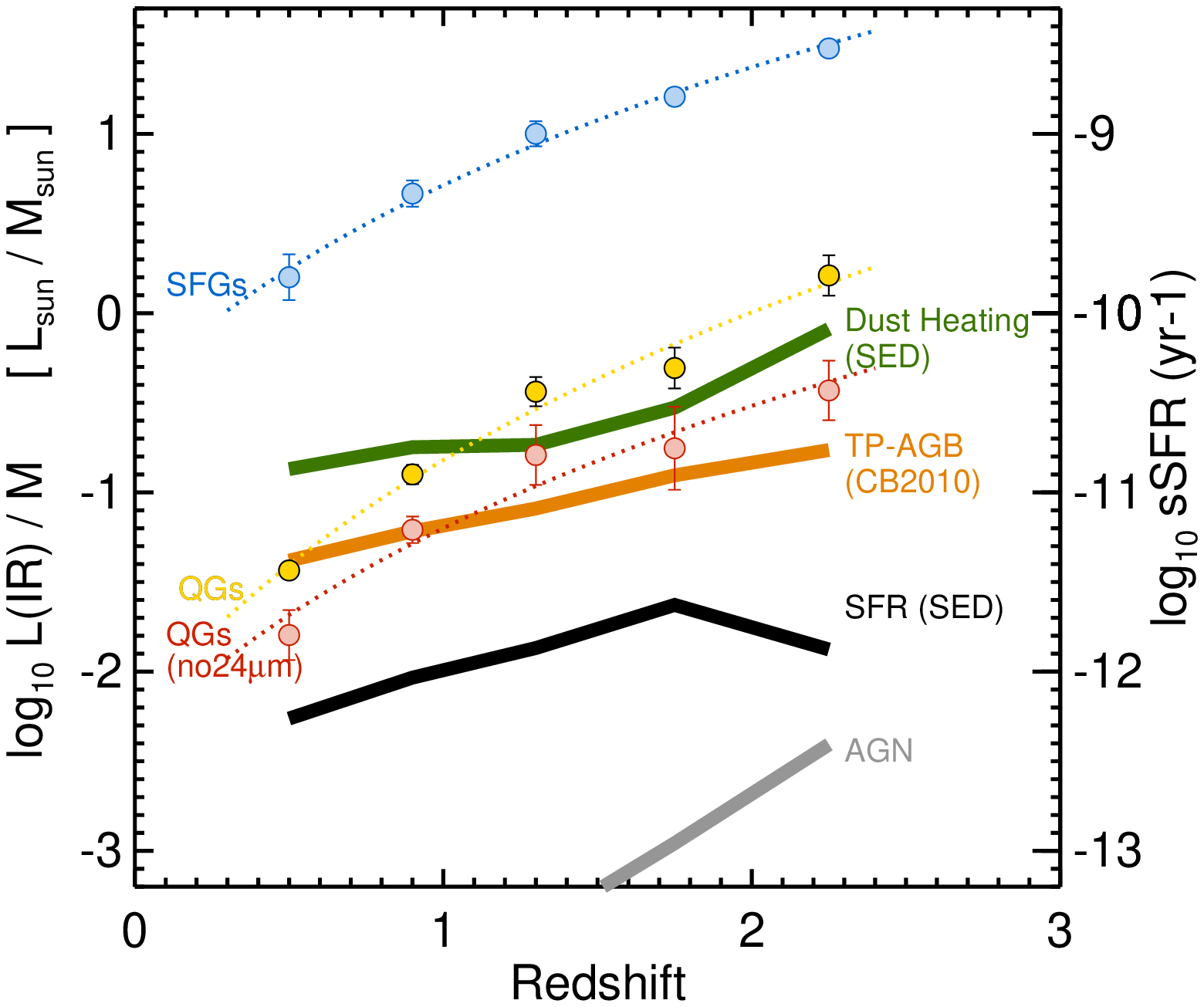}											     	  											     	  
\caption{Left: Evolution of sSFR(IR) with redshift in a ${\rm log_{10}(}M_{\star}/M_{\odot}{\rm )>10.3}$ mass-selected sample. light-blue dots indicate mean values for SFGs, while yellow and red points are stacked values of non-24$\mu$m-detected QGs (red), and all QGs (yellow). At any redshift the average QG has a sSFR 20 times lower than the star-forming sequence. The evolution of sSFR of QGs resembles the one of SFGs. As in Figure 4, we indicate with a black line the sSFR from SED fitting and with a green line the expected sSFR from mass loss. At high redshift, the sSFR(IR) of QGs is comparable to the mass-loss. Right: Comparison of observed and modeled $L_{{\rm IR}}/M_{\star}$. Values from the stacks of quiescent galaxies are represented by dotted yellow and red lines. SFGs mean values (light-blue) are put as a reference. Expected contributions to $L_{{\rm IR}}$ for the QG samples from models described in Section \ref{Contributions to LIR} are drawn with solid lines (gray: AGN, orange: circumstellar dust, black: SFR from best fits, green: cirrus dust heating). Circumstellar dust and cirrus dust can account for most of the observed $L_{{\rm IR}}$.}
\label{evolution_sSFR}
\end{figure*}

\begin{figure}[!t!h]
\centering
\includegraphics[width=8.5cm]{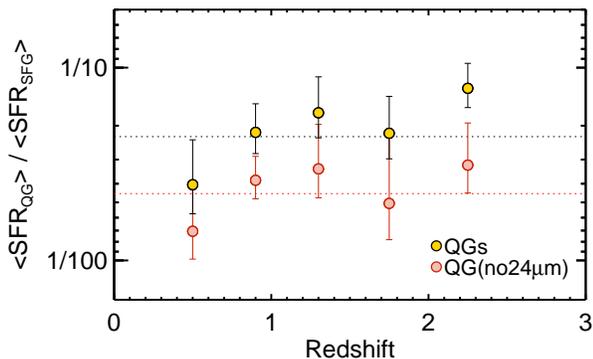}

\caption{Ratio of the average SFR of QGs to the average SFR of SFGs at the same redshift. Red dots represent QGs which are individually undetected at 24$\mu$m, while yellow dots represent all QGs. For the two samples, the average ratio is respectively $1 / (45 \pm 11)$ and $1 / (22 \pm 7)$. These ratios are possibly even lower because for QGs IR inferred SFRs can be significantly contaminated by other sources of dust heating (Section 5). }
\label{ratio_sSFR}
\end{figure}

We obtain mostly clear detections with signal-to-noise of 3--5, and fluxes $\rm F_{24 \mu m } = 2-5 \mu Jy$, corresponding to SFR $\rm \sim$ 0.5-5 $\rm M_{\odot}/yr$. We summarize the measured stacked fluxes in Table 1. Errors on the stacks are measured through bootstrapping of the sources. Errors on the stacks are measured through bootstrapping, as follows. Each sample of QGs is resampled 1000 times. We stack the individual 24$\mu$m images of galaxies belonging to each resampling and perform photometry on the new stacked images. The dispersion of the flux values in the resampled stacks gives the uncertainty on the flux measurement. 

In Figure \ref{MassSFR} we overplot with large yellow dots the SFR obtained from all QGs (big yellow dots) and non-24$\mu$m-detected galaxies (big red dots), representative of the deadest fraction of the galaxy population (this definition of 'quiescent galaxy' is the same as in Bell et al. 2012).

Despite the different sample selection (all QGs or just QGs not detected at 24$\mu$m), it is evident that at each redshift the average QGs has a SFR which is at least $\sim 20-40$ times lower than the ones on the 'star-forming sequence'.

In Figure \ref{evolution_sSFR} (left panel) we show the redshift evolution of SFRs of SFGs and QGs. We plot sSFR since it is more mildly dependent on stellar mass than SFR itself. As noted by previous studies (e.g. Damen et al. 2010, Whitaker et al. 2012, Karim et al. 2012, Fumagalli et al. 2012), the evolution of sSFR in redshift for star-forming galaxies is well fit by a power law $(1+z)^n$ where $n \sim 3-4$. At each redshift QGs have a sSFR at least 20 times lower than SFGs. The evolution with redshift of sSFR of QGs seems to resemble the evolution of SFGs. At redshift $z \sim 2$, QGs form 10 times more stars than at redshift $z \sim 0.5$. 

In Figure \ref{ratio_sSFR} we show for our samples the ratio of the average SFR of QGs to the average SFR of SFGs at the same redshift. For quiescent galaxies undetected at 24$\mu$m, the mean value of the ratio is $\rm \langle SFR_{QG} \rangle /  \langle SFR_{SFG} \rangle = 1 / (45 \pm 11)$ while for the entire sample it is $\rm \langle SFR_{QG} \rangle / \langle SFR_{SFG} \rangle = 1 / (22 \pm 7)$. This confirms that at each redshift quenching of star formation is very efficient.
 
For QGs the SFRs inferred from the IR emission are generally an order of magnitude larger than those inferred from stellar population modeling (black dashed line in Figure \ref{evolution_sSFR}, left panel). At the highest redshifts they are similar to the values predicted by the recycling of mass loss (green dashed line in Figure \ref{evolution_sSFR}, left panel), while at redshift lower than 1.5 they are significantly lower than those. 

\section{Other possible contributions to $L_{{\rm IR}}$}
\label{Contributions to LIR}

Strictly should be that the IR-inferred SFRs for QGs are upper limits, because of contributions of AGNs, AGB stars and dust heating from old stellar populations to the IR fluxes. We treat each of these components separately in the following Subsections and compare their contributions to $L_{{\rm IR}}$ with the observed stacked values of $L_{{\rm IR}}$ in the Discussion Section.

\subsection{AGN}
We evaluate the possible contribution of AGNs by stacking X-ray thumbnails (from the CDF-S 4Ms and CDF-N 2Ms) of the QGs in different redshift bins. Mullaney et al. 2011 demonstrates (Equation 4) the existence of a linear relation between the X-ray luminosity $L_{\rm X}$ and $L_{{\rm IR}}$ for a sample of local AGNs. After subtracting individually-detected X-ray point sources (marked with orange stars in Figure \ref{MassSFR}), we obtain marginal detections ($2-3 \sigma$) ranging from $L_{\rm X} {\rm \sim 3.8 \times 10^{40} erg / s}$ \footnote{X-ray luminosities are evaluated assuming a power law spectrum with $\Gamma = 1.8$} in the lowest redshift bin to $L_{\rm X} {\rm \sim 2.0 \times 10^{41} erg / s}$ in the highest redshift bin. Converting the obtained X-ray luminosities to IR luminosities with the Mullaney relation, we obtain the gray line in Figure \ref{evolution_sSFR} (right panel). It lies three orders of magnitude below the observed $\rm L_{(IR)} / M_{\star}$ \footnote{A high fraction of Compton-thick AGNs in the sample would originate a higher IR luminosity inferred from X-ray stacks. The percentage of Compton-thick AGNs is however poorly constrained at high redshift (e.g. Akylas et al. 2012).}. Olsen et al. (2013) suggest that at redshift $z \sim 2$ most QGs host a low-luminosity AGN, comparing SFR inferred from IR and X-ray. They find, for QG at $1.5 < z < 2.5$ of the same mass of that of our sample, a mean luminosity of $L_{{\rm X}} < 2.5 \times 10^{41}$, consistent with our study. Even though most of QGs host a (low-luminosity) AGN, we find that those weak AGN can not account for the MIR emission of the galaxies. Other studies (Donley et al. 2008; Kartaltepe et al. 2010) have also already pointed out that systems with 24$\mu$m flux dominated by AGNs are not the dominant population at low $L_{{\rm IR}}$, such as QGs.

\subsection{Circumstellar dust}
AGB stars are known to evolve embedded in a circumstellar dusty envelope (e.g. Bressan et al. 1998, Lancon \& Mouhcine 2002, Piovan et al. 2003). They are the dominant source of the rest-frame $K$-band luminosity between 0.1 and 1.5 Gyr of age (Kelson \& Holden 2010) and significantly contribute to MIR emission, but their dust contribution is not included in classical optical-near infrared SED fitting (Bruzual \& Charlot 2003, Maraston 2005). We evaluate the contribution to $L_{{\rm IR}}$ with the new Charlot \& Bruzual 2010 model (CB2010) of an SSP with solar metallicity (private communication). Given galaxy ages from the FAST best fits (see Section 2, and Whitaker et al. 2013, in press), for each galaxy in our QG sample we estimate the observed 24$\mu$m flux from the CB2010 model and convert it to $L_{{\rm IR}}$ with the same relation of Wuyts et al.(2008) (Figure 8, right panel, orange line). 

\subsection{Cirrus dust}
Another possible contribution to $L_{{\rm IR}}$ is dust heating from old stellar populations. Salim et al. (2009) concludes that, for a sample of 24$\mu$m-detected galaxies in the DEEP2 survey (0.2 $< z <$ 1.0), the bulk of IR emission in red ($NUV-r$) galaxies comes from the heating of diffuse cirrus dust by old stellar populations, rather than by dust heating in star-forming regions. We test if this holds true for the galaxies in our sample as follows. Given the stellar population parameters from the FAST best-fit to the SEDs (age, $\tau$, $A_{V}$), we evaluate the luminosity absorbed at $\lambda < 1 \mu$m by  integrating the difference between the unattenuated and the attenuated synthetic SED, and assume it is re-emitted in the IR (see Charlot \& Fall 2000, Da Cunha et al. 2008). 

\begin{figure}[!h!]
\centering
\includegraphics[width=8cm]{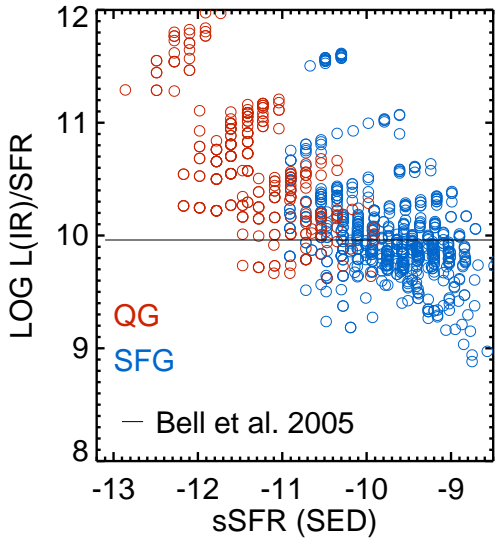}											     	  
\caption{Model predictions of $L_{{\rm IR}}$/SFR, for QGs (red) and SFGs (blue). $L_{{\rm IR}}$ is reconstructed assuming that the light absorbed by dust at UV-optical wavelengths is re-emitted in the IR (Section \ref{Contributions to LIR}). For SFGs the ratio is comparable to the Bell et al. (2005) relation (black line), while for QGs $L_{{\rm IR}}$ are systematically higher than the expectations from SFR, meaning that for QGs most of dust heating comes from old stellar populations.}
\label{predicted IR}
\end{figure}

We then compare the model $L_{{\rm IR}}$ predicted by the attenuated SED with the best fit SFR. If $L_{{\rm IR}}$ originates in dust associated with star-forming regions, we expect the ratio $L_{{\rm IR}}$/SFR to be ${\rm \sim 9.8 \times 10^{9}} L_{\odot}/M_{\odot}$ (Bell et al., 2005). Figure \ref{predicted IR} shows that SFGs (blue) are consistent with this prediction. On the other hand, for QGs (red points) $L_{{\rm IR}}$ is systematically higher than the expectations from dust heating in star-forming regions. This indicates that in QGs a significant contribution to $L_{{\rm IR}}$ comes from dust heated by old stellar populations.
Inferring SFR from $L_{{\rm IR}}$ (and therefore from 24$\mu$m fluxes) overestimates the real SFR of QGs. For each galaxy in the QG sample we estimate the expected $L_{{\rm IR}}$ luminosity from circumstellar dust, and compute the mean value in different redshift bins (Figure 8, right panel, green line).

\section{Discussion}

As we have seen above, various processes other than star formation can contribute to the
observed mid-IR flux. We next discuss the impact on the derived SFRs.
Moreover, we put constraints on the mass growth of QGs implied by the measured SFRs and on their size growth implied by the stellar mass loss.

In Figure \ref{evolution_sSFR} (right panel) we show the approximate evolution of $L_{{\rm IR}} / M_{\star}$, for data (dashed lines) and models (thick solid lines). 
We saw earlier that observations of $L_{{\rm IR}}$ are based on the extrapolation of the single band 24$\mu$m to $L$(IR) assuming a template for dust heating by star-forming regions (Section 2). Model predictions estimate that the AGN contribution (gray line) to the $L_{{\rm IR}}$ is negligible for our sample, while the model expectation for $L_{{\rm IR}}$ from cirrus dust (green) and circumstellar dust (orange) is comparable to the observed values from stacking. We note that qualitatively both of them decrease towards lower redshift, respectively because of higher $A_{V}$ and younger stellar ages at higher redshifts (which leads to more absorbed optical light re-emitted in the IR in the younger Universe) and because of the aging of galaxies (which leads to lower contribution of AGB stars in the SED).

If SFRs from SED fitting are correct, their contribution to $L_{{\rm IR}}$ (black line in Figure \ref{evolution_sSFR}, right panel) would be 1 dex lower than the observed $L_{{\rm IR}}$, while dust heated by old stellar population can account for the most of the observed luminosities. 

All the measured values from 24$\mu$m stacks must therefore be considered as upper limits to the SFR. At each redshift, the mean QG has a SFR at least $\sim$20-40 times lower than that of a SFG at the same redshift. These SFRs are significantly higher than estimates based on optical and near-IR model fits (see Section 3 and Ciambur, Kauffman \& Wuyts 2013).

In order to evaluate the growth of a QG via star formation we integrate the sSFR(IR)-z trend of Figure 8 (left). This leads to estimate that the $\it maximum$ growth of a QG via star formation is 20\% from redshift 2 to 0. Some authors (e.g. van Dokkum et al. 2010, Patel et al. 2012) have inferred that a present-day $10^{11.2}$ $\rm M_{\odot}$ galaxy has to grow 60\% of its mass from redshift $z \sim 1.75$ to $z \sim 0$. We show that star formation can not be responsible for the entire stellar growth of QGs, while other mechanisms must be in place, such as minor merging (see, among others, Hopkins 2009).
The limit we compute on the mass increase via star formation is more stringent than that computed by P{\'e}rez-Gonz{\'a}lez et al. (2008), who estimates that massive spheroid-like galaxies may have doubled (at the most) their stellar mass from redshift 2 to 0.

The SFRs expected from stellar mass loss are probably much higher than the real SFRs of QGs, meaning that star formation from mass loss is inefficient.
If mass loss from evolved stars is not converted into stars and gas is expelled from the galaxy, an interesting consequence is that the potential of the system becomes shallower and the system expands (Zhao et al. 2002, Murray et al. 2010). In brief (following Damjanov et al., 2009), if a system loses a fraction $\rm \delta M / M$ of its mass in a time longer than a dynamical timescale, it will expand its radius by a factor of $\rm \delta R / R \sim (1-\delta M / M)^{-1}$. The modeled mass losses for galaxies in our sample (Figure 4) integrated over the redshift range 0 to 2 give $\rm \delta M/M \approx 0.4$, which leads to $\rm \delta R/R \approx 0.6$. The observed size growth of quiescent galaxies from redshift 2 to 0 amounts to a factor of 2-3 (e.g. Williams et al. 2010, Newman et al. 2012, Whitaker et al. 2012), therefore mass loss can not be its unique cause but only one of the concurrent ones (see also Damjanov et al., 2009). We note that the assumed mass loss depends on the absolute ages of galaxies at each redshift, which are very uncertain.

A possible caveat in the study comes from the fact that the 24$\mu$m-to-L(IR) conversion relies on a single infrared template (Wuyts et al. 2008), while the underlying SED for high-redshift quiescent galaxies is unknown. As explained in Section 2 (Data), measuring the low fluxes of quiescent galaxies at Herschel wavelengths is extremely challenging. Available observationally motivated far-IR SEDs of galaxies at high redshift are based on bright Herschel sources (e.g. Elbaz et al. 2011, Magdis et al. 2014) and refer to galaxies on or above the star-forming main sequence. We evaluate the possible bias introduced by our synthetic template by comparing the L(IR) integrated under the Elbaz et al. 2011 SED for main-sequence galaxies to that inferred from a simulated 24$\mu$m observation of that SED, at different redshifts. We obtain that at $z>1.5$ the inferred L(IR) is a factor of 2 higher than the integrated L(IR), while at $z<1.5$ the bias is lower than 50\%. 

The possibility that galaxies below the main-sequence have different infrared SED shapes must however be taken into account (see also Utomo et al. 2014, Hayward et al. 2014). We compute the systematic uncertainty on L(IR) arising from the unknown underlying SED as follows. We compare the 24$\mu$m-to-LIR conversion of Wuyts et al. 2008, obtained by averaging a vast array of infrared templates from Dale \& Helou 2002, with those obtained by using each single Dale \& Helou 2002 template for different values of the ionization parameter $\alpha$. The dispersion on the values is 0.3 dex, which we consider the systematic uncertainty on the conversion.
We conclude that the possible biases and uncertainties induced by inferring L(IR) from a single band and a single template amount to a factor of 2, and do not affect the conclusions of the paper.

\section{Conclusions}

We select quiescent galaxies at redshift $0.3<z<2.5$ in the 3D-HST survey from their rest-frame optical and near-IR colors. Fitting their UV to near-IR photometry with stellar population models, we find very low star-formation rates ($\rm sSFR \sim 10^{-12} yr^{-1}$). These values are much lower than the stellar mass loss rates predicted by the same models. This suggests that the star formation is either missed because it is dust obscured, or that the gas from stellar mass loss is expelled from the galaxy or prevented from refueling star formation. 

We put upper limits on the obscured star-formation rate of quiescent galaxies by stacking 24$\mu$m images. Including direct 24$\mu$m detections, we find that $\rm sSFR(IR) \le 10^{-11.9}  \times (1+z)^{4} yr^{-1}$.
At each redshift the sSFR of quiescent galaxies is $\sim$ 20-40 times lower than the typical value on the main sequence of star-forming galaxies.
SFRs of quiescent galaxies are possibly even lower than this, because the IR luminosity can also be due to other sources, such as the presence of AGB dust enshrouded stars and dust heating from older stellar populations. Stacks of longer wavelength data (such as from {\it Herschel}) are necessary for constraining the dust temperature and therefore distinguishing between the different contributions to $L_{{\rm IR}}$, however a large sample may be necessary to achieve adequate S/N (e.g. Viero et al. 2013). We show nevertheless that dust heating from old stellar populations can account for most of the observed $L_{{\rm IR}}$.

The observed SFR(IR) are therefore upper limits to the real SFR, which are possibly one order of magnitude lower.
This means that there must be a mechanism that not only shuts down star formation, but also keeps the galaxy dead for a long period of time, preventing the ejected gas from cooling and forming new stars. If gas from mass-loss is expelled from galaxies, we predict that it is responsible for a growth in stellar radii of 60\% from redshift 2 to 0.

\acknowledgments 
We acknowledge funding from ERC grant HIGHZ no. 227749. This work is based on observations taken by the 3D-HST Treasury Program (GO 12177 and 12328) with the NASA/ESA HST, which is operated by the Association of Universities for Research in Astronomy, Inc., under NASA contract NAS5-26555.

\begin{table*}[!h!]
  \caption{Properties of Stacks}
  \label{Stacks}
  \begin{center}

    \leavevmode
\begin{tabular}{l|lll|lll} \hline \hline            
 Redshift    &       $\rm N_{QG}$       &      $\rm F(24 \mu m) _{QG}$    &  $\rm SFR(IR)_{QG}$  &  $\rm N_{QG, no24\mu m}$ &  $\rm F(24 \mu m) _{QG, no24\mu m}$   &  $\rm SFR(IR)_{QG, no24\mu m}$ \\ \hline 
 
 0.3 - 0.7 & 97 & $\rm  3.9 \pm  1.3 \ \mu Jy$ &  $\rm 0.2 \pm   0.1\ M_{\odot} / yr $&67 & $\rm  7.7 \pm  0.5\ \mu Jy$ &  $\rm 0.4 \pm   0.1\ M_{\odot} / yr$ \\
 0.7 - 1.1 & 154 & $\rm  4.1 \pm  0.7\ \mu Jy$ &  $\rm 0.5 \pm   0.1\ M_{\odot} / yr $&108 & $\rm  6.6 \pm  0.7\ \mu Jy$ &  $\rm 1.2 \pm   0.1\ M_{\odot} / yr$ \\
 1.1 - 1.5 & 84 & $\rm  4.4 \pm  1.8\ \mu Jy$ &  $\rm 1.8 \pm   0.6\ M_{\odot} / yr $&58 & $\rm  9.3 \pm  1.8\ \mu Jy$ &  $\rm 3.7 \pm   0.7\ M_{\odot} / yr$ \\
 1.5 - 2.0 & 72 & $\rm  3.0 \pm  1.6\ \mu Jy$ &  $\rm 2.0 \pm   1.0\ M_{\odot} / yr $&51 & $\rm  6.8 \pm  1.8\ \mu Jy$ &  $\rm 4.6 \pm   1.3\ M_{\odot} / yr$ \\
 2.0 - 2.5 & 35 & $\rm  3.2 \pm  1.3\ \mu Jy$ &  $\rm 3.8 \pm   1.5\ M_{\odot} / yr $&25 & $\rm  5.7 \pm  1.8\ \mu Jy$ &  $\rm 8.8 \pm   2.2\ M_{\odot} / yr$ \\

\hline

\end{tabular}
\bigskip

For different redshift bins: number of galaxies in the quiescent sample (QG) and quiescent sample without 24$\mu$m detection (QG,no24$\mu$m), along with their stacked 24m fluxes, and the implied SFR from IR emission.

\end{center}
\end{table*}

\clearpage

\appendix
\label{Appendix}
\section{A. Photometry}

\setcounter{figure}{0} \renewcommand{\thefigure}{A\arabic{figure}}

The MIPS-24$\mu$m beam has a FWHM of 6 arcsec, therefore confusion and blending effects are unavoidable in deep observations at this resolution. 
We use a source-fitting algorithm designed to extract photometry from IRAC and MIPS images (see, e.g., Labbe ́ et al. 2006; Wuyts et al. 2007). The information on position and extent of the sources based on the higher resolution $F160W$ segmentation map is used to model the lower resolution  MIPS-24$\mu$m images. Local convolution kernels are constructed using bright, isolated, and unsaturated sources in the $F160W$ and MIPS-24$\mu$m, derived by fitting a series of Gaussian-weighted Hermite functions to the Fourier transform of the sources. 
Each source is extracted separately from the $F160W$ image and, under the assumption of negligible morphological K-corrections, convolved to the MIPS-24$\mu$m resolution using the local kernel coefficients.
All sources in each MIPS-24$\mu$m image are fit simultaneously, with the flux left as the only free parameter. The modeled light of neighboring sources (closer than 10 arcsec) is subtracted, thereby leaving a "clean" MIPS-24$\mu$m image to perform aperture photometry and stacking of faint sources. The technique is illustrated in Figure A1 and A2, respectively for a bright and a faint source.

\begin{figure}[!h!]
\centering
\includegraphics[width=10cm]{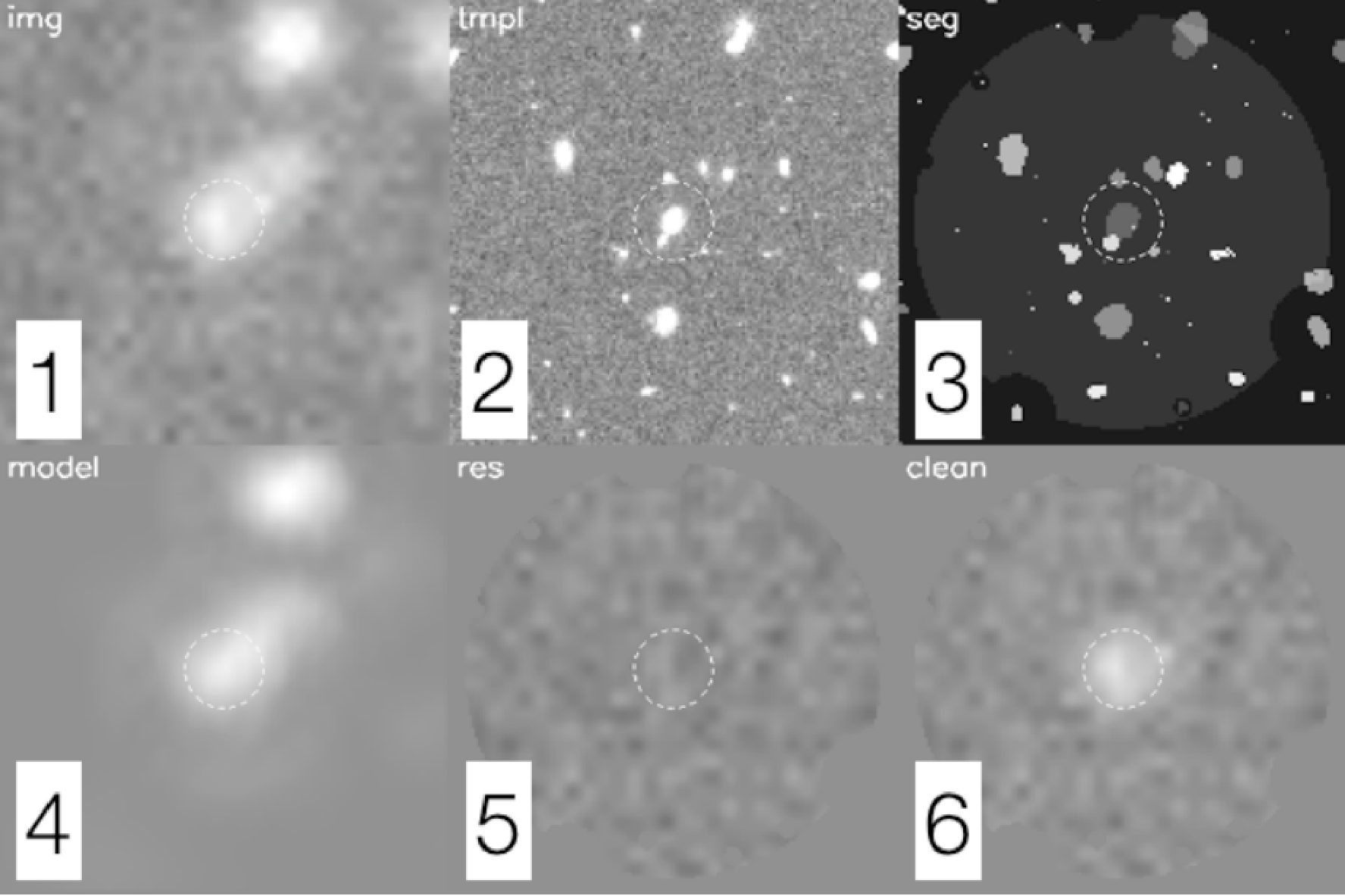}											     	  
\caption{The process of modeling and deblending 24$\mu$m fluxes for objects identified in the F160W detection image. Panel 1 shows the original 24$\mu$m cutout for an object in the catalog. Panel 2 and 3 show the matching F160W detection image and segmentation map from SExtractor. The bottom row shows the modeled 24$\mu$m flux for all objects in the region (Panel 4), the residual image with all modeled fluxes removed (Panel 5), and the flux for the central object alone (Panel 6).}
\label{Figure A1}
\end{figure}

\begin{figure}[!h!]
\centering
\includegraphics[width=10cm]{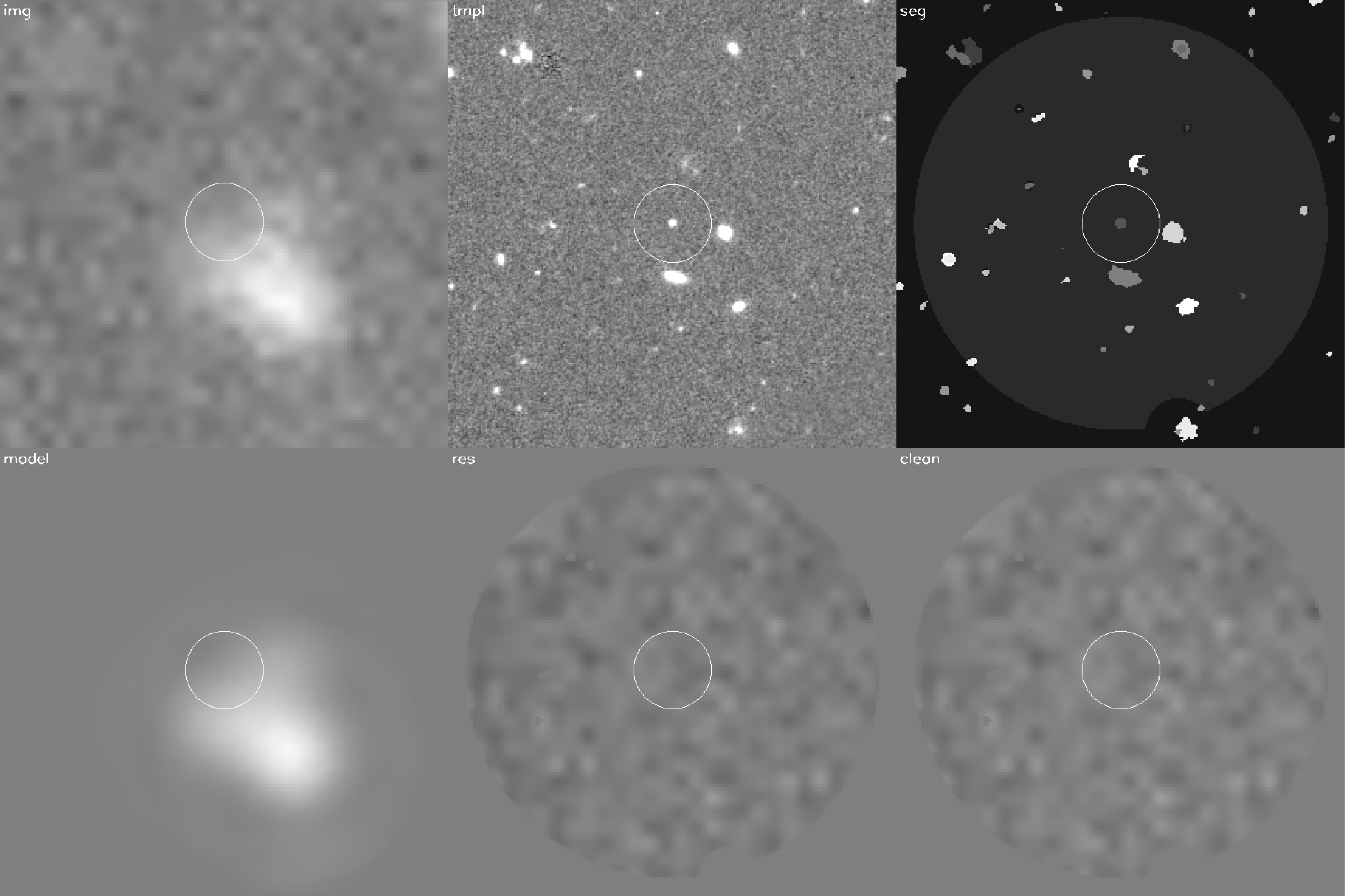}											     	  
\caption{Same as Figure A1, but for a faint object in the catalog.}
\label{Figure A2}
\end{figure}

\section{B. Field-to-field variation}
\setcounter{figure}{0} \renewcommand{\thefigure}{B\arabic{figure}}

The paper is built on data from the GOODS-North and GOODS-South fields. The two fields feature a similar large number of optical-near-IR observations included in the 3D-HST photometric catalog, and data quality in the 3D-HST fields is uniform (see Skelton et al., 2014). The depths of MIPS-24$\mu$m data are similar in the two fields (Dickinson et al. 2003). We show in Figure B1 the main result of the paper - the evolution of sSFRs of QGs - once the data are stacked separately in the two fields. Differences and errors are consistent with lower statistics.

\begin{figure}[!h!]
\centering
\includegraphics[width=8.5cm]{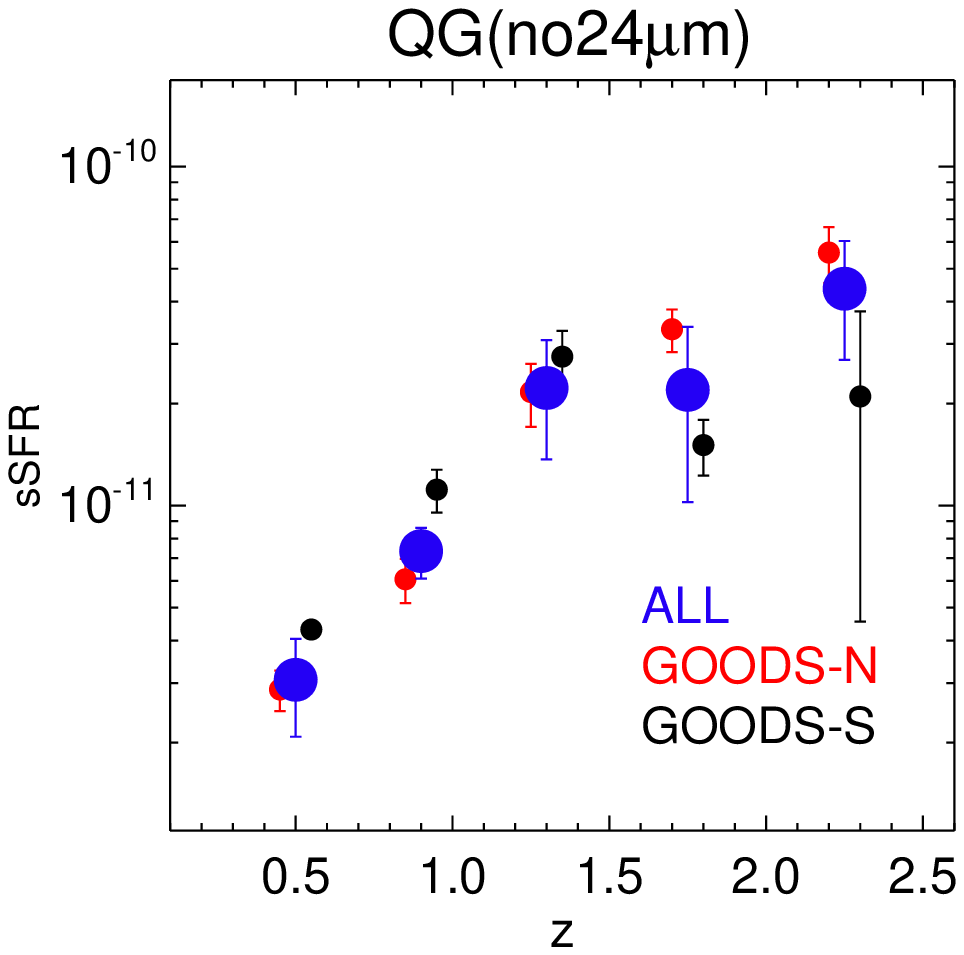}											     	  
\caption{Evolution of sSFR of QGs without an individual 24$\mu$m detection, for different fields and in the entire sample. Red/black dots are stacked values from GOODS-North/GOODS-South, and large blue dots are values from the combined sample. Errors are computed bootstrapping the sample. Mean redshifts have been shifted of $\pm$ 0.05 for clarity. Differences and errors are consistent with lower statistics.}
\label{Figure B1}
\end{figure}


\begin{thebibliography}{}

\bibitem[Akylas et 
al.(2012)]{2012A&A...546A..98A} Akylas, A., Georgakakis, A., Georgantopoulos, I., Brightman, M., \& Nandra, K.\ 2012, \aap, 546, A98 

\bibitem[Alexander et al.(2003)]{2003AJ....126..539A} Alexander, D.~M., 
Bauer, F.~E., Brandt, W.~N., et al.\ 2003, \aj, 126, 539 

\bibitem[Barro et al.(2013)]{2013ApJ...765..104B} Barro, G., Faber, S.~M., 
P{\'e}rez-Gonz{\'a}lez, P.~G., et al.\ 2013, \apj, 765, 104 

\bibitem[Bell et al.(2005)]{2005ApJ...625...23B} Bell, E.~F., Papovich, C., 
Wolf, C., et al.\ 2005, \apj, 625, 23 

\bibitem[Bell et al.(2012)]{2012ApJ...753..167B} Bell, E.~F., van der Wel, 
A., Papovich, C., et al.\ 2012, \apj, 753, 167 

\bibitem[Brammer et al.(2008)]{2008ApJ...686.1503B} Brammer, G.~B., van 
Dokkum, P.~G., \& Coppi, P.\ 2008, \apj, 686, 1503 

\bibitem[Brammer et al.(2009)]{2009ApJ...706L.173B} Brammer, G.~B., 
Whitaker, K.~E., van Dokkum, P.~G., et al.\ 2009, \apjl, 706, L173 

\bibitem[Brammer et al.(2011)]{2011ApJ...739...24B} Brammer, G.~B., 
Whitaker, K.~E., van Dokkum, P.~G., et al.\ 2011, \apj, 739, 24 

\bibitem[Brammer et al.(2012)]{2012arXiv1204.2829B} Brammer, G., van 
Dokkum, P., Franx, M., et al.\ 2012, arXiv:1204.2829 

\bibitem[Parriott 
\& Bregman(2008)]{2008ApJ...681.1215P} Parriott, J.~R., \& Bregman, J.~N.\ 2008, \apj, 681, 1215 


\bibitem[Bressan et 
al.(1998)]{1998A&A...332..135B} Bressan, A., Granato, G.~L., \& Silva, L.\ 1998, \aap, 332, 135 

\bibitem[Bruzual 
\& Charlot(2003)]{2003MNRAS.344.1000B} Bruzual, G., \& Charlot, S.\ 2003, \mnras, 344, 1000 

\bibitem[Chabrier(2003)]{2003ApJ...586L.133C} Chabrier, G.\ 2003, \apjl, 
586, L133 

\bibitem[Charlot 
\& Fall(2000)]{2000ApJ...539..718C} Charlot, S., \& Fall, S.~M.\ 2000, \apj, 539, 718 

\bibitem[Ciambur et al.(2013)]{2013MNRAS.432.2488C} Ciambur, B.~C., 
Kauffmann, G., \& Wuyts, S.\ 2013, \mnras, 432, 2488 

\bibitem[da Cunha et al.(2008)]{2008MNRAS.388.1595D} da Cunha, E., Charlot, 
S., \& Elbaz, D.\ 2008, \mnras, 388, 1595 

\bibitem[Damen et al.(2011)]{2011ApJ...727....1D} Damen, M., Labb{\'e}, I., 
van Dokkum, P.~G., et al.\ 2011, \apj, 727, 1 

\bibitem[Damjanov et al.(2009)]{2009ApJ...695..101D} Damjanov, I., 
McCarthy, P.~J., Abraham, R.~G., et al.\ 2009, \apj, 695, 101 

\bibitem[Dickinson 
\& GOODS Team(2004)]{2004AAS...20516301D} Dickinson, M., \& GOODS Team 2004, Bulletin of the American Astronomical Society, 36, \#163.01 

\bibitem[Donley et al.(2008)]{2008ApJ...687..111D} Donley, J.~L., Rieke, 
G.~H., P{\'e}rez-Gonz{\'a}lez, P.~G., \& Barro, G.\ 2008, \apj, 687, 111 

\bibitem[Elbaz et 
al.(2011)]{2011A&A...533A.119E} Elbaz, D., Dickinson, M., Hwang, H.~S., et al.\ 2011, \aap, 533, A119 



\bibitem[Franx et al.(2008)]{2008ApJ...688..770F} Franx, M., van Dokkum, 
P.~G., Schreiber, N.~M.~F., et al.\ 2008, \apj, 688, 770 

\bibitem[Fumagalli et al.(2012)]{2012ApJ...757L..22F} Fumagalli, M., Patel, 
S.~G., Franx, M., et al.\ 2012, \apjl, 757, L22 

\bibitem[Gobat et al.(2013)]{2013arXiv1305.3576G} Gobat, R., Strazzullo, 
V., Daddi, E., et al.\ 2013, arXiv:1305.3576 

\bibitem[Hopkins et al.(2009)]{2009MNRAS.398..898H} Hopkins, P.~F., Bundy, 
K., Murray, N., et al.\ 2009, \mnras, 398, 898 

\bibitem[Ilbert et al.(2010)]{2010ApJ...709..644I} Ilbert, O., Salvato, M., 
Le Floc'h, E., et al.\ 2010, \apj, 709, 644 

\bibitem[Karim et al.(2011)]{2011ApJ...730...61K} Karim, A., Schinnerer, 
E., Mart{\'{\i}}nez-Sansigre, A., et al.\ 2011, \apj, 730, 61 

\bibitem[Kartaltepe et al.(2010)]{2010ApJ...709..572K} Kartaltepe, J.~S., 
Sanders, D.~B., Le Floc'h, E., et al.\ 2010, \apj, 709, 572 

\bibitem[Kauffmann et al.(2003)]{2003MNRAS.341...54K} Kauffmann, G., 
Heckman, T.~M., White, S.~D.~M., et al.\ 2003, \mnras, 341, 54 

\bibitem[Kelson 
\& Holden(2010)]{2010ApJ...713L..28K} Kelson, D.~D., \& Holden, B.~P.\ 2010, \apjl, 713, L28 

\bibitem[Kennicutt(1998)]{1998ApJ...498..541K} Kennicutt, R.~C., Jr.\ 1998, 
\apj, 498, 541 

\bibitem[Komatsu et al.(2011)]{2011ApJS..192...18K} Komatsu, E., Smith, 
K.~M., Dunkley, J., et al.\ 2011, \apjs, 192, 18 

\bibitem[Kriek et al.(2006)]{2006ApJ...649L..71K} Kriek, M., van Dokkum, 
P.~G., Franx, M., et al.\ 2006, \apjl, 649, L71 

\bibitem[Kriek et al.(2009)]{2009ApJ...700..221K} Kriek, M., van Dokkum, 
P.~G., Labb{\'e}, I., et al.\ 2009, \apj, 700, 221 

\bibitem[Kriek et al.(2011)]{2011ApJ...743..168K} Kriek, M., van Dokkum, 
P.~G., Whitaker, K.~E., et al.\ 2011, \apj, 743, 168 


\bibitem[Labb{\'e} et al.(2005)]{2005ApJ...624L..81L} Labb{\'e}, I., Huang, 
J., Franx, M., et al.\ 2005, \apjl, 624, L81 

\bibitem[Labb{\'e} et al.(2006)]{2006ApJ...649L..67L} Labb{\'e}, I., 
Bouwens, R., Illingworth, G.~D., \& Franx, M.\ 2006, \apjl, 649, L67 


\bibitem[Lan{\c c}on 
\& Mouhcine(2002)]{2002A&A...393..167L} Lan{\c c}on, A., \& Mouhcine, M.\ 2002, \aap, 393, 167 

\bibitem[Leitner 
\& Kravtsov(2011)]{2011ApJ...734...48L} Leitner, S.~N., \& Kravtsov, A.~V.\ 2011, \apj, 734, 48 

\bibitem[Maraston(2005)]{2005MNRAS.362..799M} Maraston, C.\ 2005, \mnras, 
362, 799 

\bibitem[Mullaney et al.(2011)]{2011MNRAS.414.1082M} Mullaney, J.~R., 
Alexander, D.~M., Goulding, A.~D., 
\& Hickox, R.~C.\ 2011, \mnras, 414, 1082 

\bibitem[Muzzin et al.(2010)]{2010ApJ...725..742M} Muzzin, A., van Dokkum, 
P., Kriek, M., et al.\ 2010, \apj, 725, 742 

\bibitem[Murray et al.(2010)]{2010ApJ...709..191M} Murray, N., Quataert, 
E., \& Thompson, T.~A.\ 2010, \apj, 709, 191 

\bibitem[Newman et al.(2012)]{2012ApJ...746..162N} Newman, A.~B., Ellis, 
R.~S., Bundy, K., \& Treu, T.\ 2012, \apj, 746, 162 

\bibitem[Noeske et al.(2007)]{2007ApJ...660L..43N} Noeske, K.~G., Weiner, 
B.~J., Faber, S.~M., et al.\ 2007, \apjl, 660, L43 

\bibitem[Olsen et al.(2013)]{2013ApJ...764....4O} Olsen, K.~P., Rasmussen, 
J., Toft, S., \& Zirm, A.~W.\ 2013, \apj, 764, 4 

\bibitem[Patel et al.(2012)]{2012ApJ...748L..27P} Patel, S.~G., Holden, 
B.~P., Kelson, D.~D., et al.\ 2012, \apjl, 748, L27 

\bibitem[P{\'e}rez-Gonz{\'a}lez et al.(2008)]{2008ApJ...687...50P} 
P{\'e}rez-Gonz{\'a}lez, P.~G., Trujillo, I., Barro, G., et al.\ 2008, \apj, 
687, 50 


\bibitem[Piovan et 
al.(2003)]{2003A&A...408..559P} Piovan, L., Tantalo, R., \& Chiosi, C.\ 2003, \aap, 408, 559 

\bibitem[Rudnick et al.(2003)]{2003ApJ...599..847R} Rudnick, G., Rix, 
H.-W., Franx, M., et al.\ 2003, \apj, 599, 847 

\bibitem[Salim et al.(2009)]{2009ApJ...700..161S} Salim, S., Dickinson, M., 
Michael Rich, R., et al.\ 2009, \apj, 700, 161 

\bibitem[Skelton et al.(2014)]{2014arXiv1403.3689S} Skelton, R.~E., 
Whitaker, K.~E., Momcheva, I.~G., et al.\ 2014, arXiv:1403.3689 

\bibitem[Sturm et 
al.(1996)]{1996A&A...315L.133S} Sturm, E., Lutz, D., Genzel, R., et al.\ 1996, \aap, 315, L133 

\bibitem[Taylor et al.(2009)]{2009ApJS..183..295T} Taylor, E.~N., Franx, 
M., van Dokkum, P.~G., et al.\ 2009, \apjs, 183, 295 


\bibitem[van Dokkum et al.(2010)]{2010ApJ...709.1018V} van Dokkum, P.~G., 
Whitaker, K.~E., Brammer, G., et al.\ 2010, \apj, 709, 1018 

\bibitem[Viero et al.(2013)]{2013arXiv1304.0446V} Viero, M.~P., Moncelsi, 
L., Quadri, R.~F., et al.\ 2013, arXiv:1304.0446 


\bibitem[Xue et al.(2011)]{2011ApJS..195...10X} Xue, Y.~Q., Luo, B., 
Brandt, W.~N., et al.\ 2011, \apjs, 195, 10 

\bibitem[Whitaker et al.(2010)]{2010ApJ...719.1715W} Whitaker, K.~E., van 
Dokkum, P.~G., Brammer, G., et al.\ 2010, \apj, 719, 1715 

\bibitem[Whitaker et al.(2012)]{2012ApJ...754L..29W} Whitaker, K.~E., van 
Dokkum, P.~G., Brammer, G., \& Franx, M.\ 2012, \apjl, 754, L29 

\bibitem[Whitaker et al.(2013)]{2013ApJ...770L..39W} Whitaker, K.~E., van 
Dokkum, P.~G., Brammer, G., et al.\ 2013, \apjl, 770, L39 

\bibitem[Williams et al.(2009)]{2009ApJ...691.1879W} Williams, R.~J., 
Quadri, R.~F., Franx, M., van Dokkum, P., 
\& Labb{\'e}, I.\ 2009, \apj, 691, 1879 

\bibitem[Wuyts et al.(2007)]{2007ApJ...655...51W} Wuyts, S., Labb{\'e}, I., 
Franx, M., et al.\ 2007, \apj, 655, 51 

\bibitem[Wuyts et al.(2008)]{2008ApJ...682..985W} Wuyts, S., Labb{\'e}, I., 
Schreiber, N.~M.~F., et al.\ 2008, \apj, 682, 985 

\bibitem[Wuyts et al.(2011)]{2011ApJ...738..106W} Wuyts, S., F{\"o}rster 
Schreiber, N.~M., Lutz, D., et al.\ 2011, \apj, 738, 106 

\end{thebibliography}
\end{document}